\documentclass[prl,amsmath,aps,superscriptaddress,twocolumn]{revtex4}
\usepackage{graphicx,xspace}
\usepackage{float}
\usepackage{amsmath}
\usepackage{amssymb}
\usepackage{color,soul}
\usepackage{epsfig}
\usepackage{graphicx,psfrag,xspace}

\begin{document}

\title{A model of dispersive transport across sharp interfaces between porous materials}
\author{Andrea Cortis}
\email{acortis@lbl.gov}
\affiliation{Earth Sciences Division, Lawrence Berkeley National Laboratory, Berkeley, California 94720, USA}
\author{Andrea Zoia}
\affiliation{CEA/Saclay, DEN/DM2S/SFME/LSET, B\^at.~454, 91191 Gif-sur-Yvette Cedex, France}

\begin{abstract}
Recent laboratory experiments on solute migration in composite porous columns (Berkowitz et al.~\cite{cortis_wrr}) have shown an asymmetry in the solute arrival time upon reversal of the flow direction, which is not explained by current paradigms of transport. In this work, we propose a definition for the solute flux across sharp interfaces and explore the underlying microscopic particle dynamics by applying Monte Carlo simulation. Our results are consistent with the findings reported in~\cite{cortis_wrr} and explain the observed transport asymmetry. An interpretation of the proposed physical mechanism in terms of a flux rectification is also provided. The approach is quite general and can be extended to other situations involving transport across sharp interfaces.
\end{abstract}
\maketitle

Berkowitz et al.~\cite{cortis_wrr} have recently reported experimental results for solute transport across the sharp interface between two porous materials, which show that, contrary to current model predictions, solute arrival times depend on the direction of the flow crossing the sharp interface. Understanding the precise physics of particle transport across sharp interfaces is of critical importance also for such diverse applications as heat flow in composite media~\cite{barrat}, or the pattern evolution of morphogens governing living tissue development~\cite[]{hornung}. The aim of this work is to provide a consistent physical model for the flux of a solute across two porous materials with different dispersive properties that explains the experimental results presented in~\cite{cortis_wrr}.

Transport processes can be conveniently described in terms of particles stochastically traveling within a porous medium. Depending on the intensity of the underlying velocity field, transport in porous materials may assume a dispersive or a diffusion-dominated character, in both cases leading to significant spread of an initially close ensemble of particles. The time evolution of this spreading is governed by a simple mass-balance equation that relates the time derivative of the particles concentration $c(x,t)$ to the divergence of the particles flux $j(x,t)$, $\partial_t c(x,t) = -\nabla\cdot j(x,t)$. The interaction of the tracer particles with the fluid flow manifests itself at the {\em macroscopic} scale as a complex combination of diffusion, dispersion, and time-memory effects, which are ultimately determined by the details of the pore geometry~\cite[]{rev_geo}.

Transport across an interface is a deceptively simple problem that has long been studied~\cite{vangenuchten2, novakowski, schwartz, leij}, and for which several alternative models have been proposed. Some of these models assume
equality of fluxes and concentrations at the interface~\cite[]{crank, labolle1}. Other studies relax the assumption on equality of concentrations, but either neglect advection~\cite{hornung}, or consider only concentration profiles~\cite{zoia1, zoia2}.
None of these models has been so far corroborated by experimental evidence of a concentration jump at the interface.

To fix the ideas, let a composite porous column of length $L=1$ consist of two half-columns of diameter $D$, filled with `fine' (F) and `coarse' (C) random arrangements of spheres of diameter $d_F$ and $d_C$, respectively, with $d_F < d_C$. A fluid flux, $Q$, can be injected from either end, fine or coarse, of the column. We assume that the aspect ratio of the column, $L/D$, is sufficiently large to justify a one-dimensional treatment. NMR experiments show that, for random packings of spheres, porosity $n$, electrical tortuosity, and diffusivity do not depend on the spheres' diameter, $d$~\cite[]{mair}. From the fluid incompressibility condition, the mean transport velocity $v = Q/(n \pi D^2/4)$ also does not depend on $d$. Moreover, for the P{\'e}clet numbers considered in~\cite[]{cortis_wrr}, $Pe=[0.3-10]$, diffusion is negligible with respect to dispersion.
The relevant material properties that depend on the sphere diameter $d$ are the permeability $k \sim d^2$, which characterizes the fluid flow in the pore geometry, and the dispersivity length $\alpha$,
the ratio of the effective dispersion coefficient $D$ in the Taylor-Aris regime to the pore velocity $v$,
which characterizes the {\em macroscopic} spreading of the tracer plume through the {\em microscopic} pore geometry, $\alpha \equiv D/v \sim d$  \cite{cortis_chen,rev_geo}.

While each layer can separately be regarded as being homogeneous, we have now a sharp interface (i.e., a macroscopic heterogeneity) at $x = 1/2$, the junction between the layers. With reference to the fine-to-coarse (F$\to$C) direction, we assume that $\alpha(x)$ can be represented by a Heaviside step function, i.e., $\alpha(x)$ takes the values $\alpha_F$, $\alpha_C$, and $\bar{\alpha}= (\alpha_F+\alpha_C)/2$, for $x<1/2$, $x>1/2$, and $x=1/2$, respectively. An analogous variation will be assumed for the the permeability $k_F< k(x) < k_C$. Obvious modifications hold for the coarse-to-fine (C$\to$F) flow direction.

Concerning the particles flux, it is customary to adopt a Fickian advection-dispersion (AD) constitutive relationship \mbox{$j^{AD}(x,t) \equiv v\left[c(x,t) - \alpha(x)\partial_x c(x,t)\right]$} [9]. The AD model predicts no difference in the
arrival times at the column outlet, as readily shown by numerical integration. This is
however inconsistent with the experimental evidence \cite{cortis_wrr}, where the solute transported
in the F$\to$C direction is observed to arrive faster with respect to the one transported
in the C$\to$F direction. Similarly, the adoption of a Fokker-Planck (FP) constitutive relationship
\mbox{$j^{FP}(x,t) \equiv v\left[(1 - \partial_x \alpha(x))c(x,t) - \alpha(x)\partial_x c(x,t) \right]$} as in [14] predicts faster solute arrival times are for the C$\to$F flow direction and thus does not explain the results in \cite{cortis_wrr}.

\begin{figure}[t]
\centerline{\epsfxsize=7.0cm\epsfbox{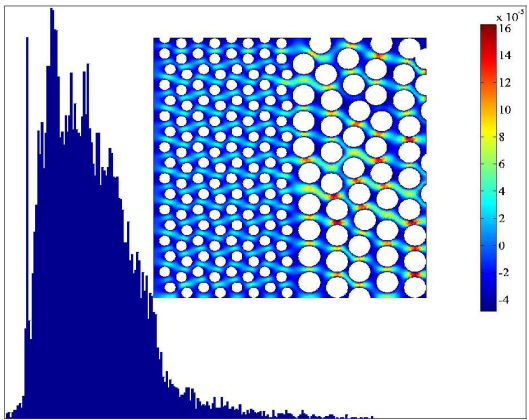}}
\caption{Deviations from the average Stokes velocity, \mbox{$u'=u-<u>$}, at the interface between two-dimensional disks (arbitrary units). The histogram of $u'$ for the F$\to$C flow direction exhibits a distinct positive skewness. Negative skewness is observed for the C$\to$F direction (not shown).}
   \label{fig:stokes}
\end{figure}

Bearing in mind these results, in the following we propose a model for the microscopic dynamics of a passive tracer across the sharp interface between two otherwise homogeneous porous materials. To this end, we begin by resorting to the Continuous Time Random Walk (CTRW) formalism, representing the stochastic path of each particle as a series of random jumps between spatial sites, separated by random waiting times. We denote the jump-length probability density function (pdf) by $p(s)$ and the waiting time pdf by $\psi(t)$, and we assume for the sake of simplicity that the two pdfs are uncoupled. The pdf $\psi(t)$ plays a central role in CTRW, since it defines the variability of the velocity spectrum and thus condenses the degree of heterogeneity of the traversed medium~\cite[]{rev_geo}. As a first approximation, the two columns can be separately considered as being homogeneous and can thus be described by a narrow $\psi(t)$ distribution (e.g., an exponential pdf), so that a single time scale $\tau$ (e.g., the mean of the distribution) dominates, and particle sojourns at each spatial site have on average the same duration. Since the two layers have the same porosity, $n$, we also assume that the time scale $\tau$ is the same in the two layers.

\begin{figure}[t]
\centerline{\epsfxsize=8.0cm\epsfbox{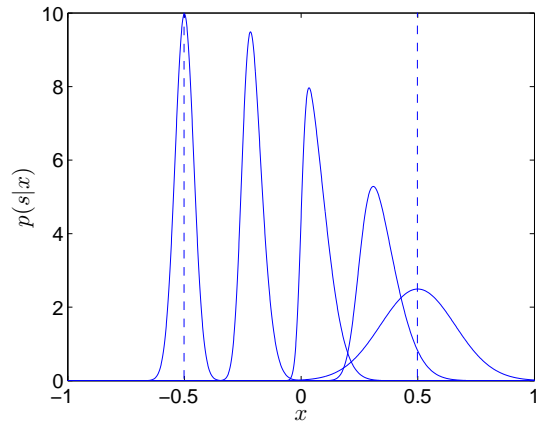}}
\caption{Skew-normal jump-length pdfs $p(s|x)$ for the case of F$\to$C flow direction. The transition region is delimited by two vertical dashed lines.}
   \label{fig:skewed_pdfs}
\end{figure}

The flux expression $j^{AD}$ is based on the assumption that the jump lengths are distributed according to a Gaussian pdf, \mbox{$g(s) \equiv \frac{1}{\sigma\sqrt{2 \pi}}e^{-(s-\mu)^2/2\sigma^2}$}~\cite{rev_geo}. The mean value of $g(s)$, $\mu$, characterizes the advective contribution to the flow, whereas $\sigma$, the square root of the variance, characterizes the dispersion, $\sigma \equiv \sqrt{2 \alpha \mu}$. The presence of the macroscopic interface imposes a spatial variability on the value of $\sigma$, but not on the value of $\mu$ because the porosity in the two sections is the same, so that we can make the identification, $p(s|x) \equiv g(s|\mu,\sigma(x))$. This pdf reproduces the Fickian flux $j^{AD}$, and consistently yields identical solute arrival times at the outlet upon reversal of the flow direction.
%

This classical picture of transport needs now to be modified to account for the local effects of the interface reported in \cite{cortis_wrr}. Consider a Lagrangian coordinate frame moving with velocity $v=\mu/\tau$. A particle located in the fine (coarse) homogeneous material at a sufficiently large distance from the interface must have an identical probability of jumping in either direction: in this region, $p(s|x)$ is symmetrical and entirely characterized in terms of $v$ and (either) $\alpha_F$ (fine layer) or $\alpha_C$ (coarse layer).
In the transition region $\mathcal{I}$, however, a particle has a finite probability of starting and ending the jump in two distinct layers, and will thus be subject to an asymmetric microscopic random force field.
For the case of F$\to$C flow direction, the probability of jumping forward will be larger than the probability of jumping backwards. Such a particle will thus locally experience a forward drift, i.e., a positively skewed jump length distribution. The same reasoning holds, but with opposite signs, for a particle crossing the interface in the C$\to$F direction; such a particle will experience a backward drift, i.e., a negatively skewed jump length distribution.
Skewed Brownian motion schemes have been proposed when the forward or backward direction of the jumps can be characterized by a biased Poisson process (see, e.g.,~\cite[]{Lejay}), and for flow parallel to a sharp interface (see, e.g.,~\cite[]{ramirez}).

\begin{figure}[t]
\centerline{\epsfxsize=9.0cm\epsfbox{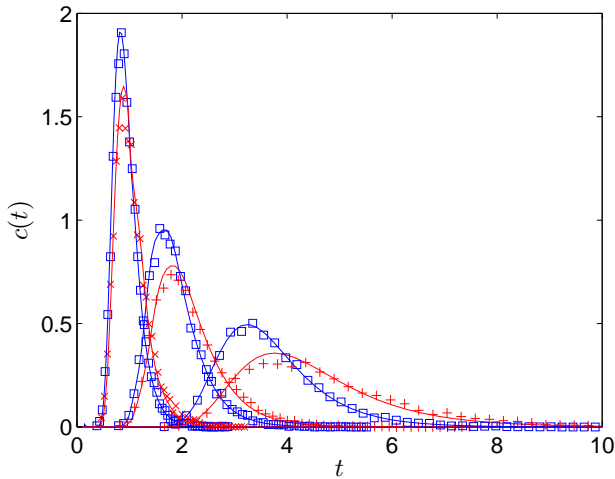}}
\caption{Breakthrough curves (BTCs) corresponding to three values of velocity $v=0.25$, $0.5$ and $1$. Dispersivities are $\alpha_F=0.1$ and $\alpha_C=0.8$, and $\bar{\theta}=0.35$. Monte Carlo simulation results are displayed as symbols: squares correspond to F$\to$C flow conditions, crosses to C$\to$F. Comparisons with the rectified flux model are shown as solid lines. For the sake of clarity, the range of Peclet numbers used in these simulations is roughly ten times larger than for the experiments in \cite{cortis_wrr}.}
   \label{fig:btc_drift}
\end{figure}

The microscopic nature of the jump length pdf asymmetry can be appreciated by observing the behavior of the pore-scale Stokes flow velocity field, $u$, across the sharp interface. It is well known from homogenization theory that in homogeneous porous media deviations of the Stokes velocity from its mean, $u' = u - \langle u \rangle$, are proportional to the permeability $k$~\cite{Sanchez}. Across the interface of a composite medium we have therefore that $|u'_C|>|u'_F|$, so that the $u'$ pdf must be skewed. Figure~\ref{fig:stokes} shows a simple 2D Stokes flow velocity field finite elements simulation~\cite{comsol} at the transition between fine and coarse disks arrangements: the histogram of the $u'$ component along the flow direction is positively (negatively) skewed in the F$\to$C (C$\to$F) flow direction.

A simple, yet realistic way to incorporate this behavior into the functional form of $p(s|x\in \mathcal{I})$ is to assume that the shape of $p(s|x)$ gradually changes its skewness within the region $\mathcal{I}$, in such a way that the skewness vanishes when particles are sufficiently far from the interface.
%
%
An expedient pdf that reproduces these features is the skew-normal distribution~\cite{azzalini}, \mbox{$f(s) \equiv \frac{1}{\sigma\pi}e^{-(s-\mu)^2/2\sigma^2} \int_{-\infty}^{\theta (s-\mu)/\sigma} e^{-y^2/2} dy$},
where $\theta$ determines the skewness of the pdf~\cite{azzalini}. Positive (negative) values of $\theta$ correspond to right (left) skewed pdfs, whereas for $\theta=0$ one recovers the normal (symmetrical) pdf, $f(s|\theta=0)=g(s)$.
The mean of $f(s)$ equals $\langle s \rangle = \mu + \xi$, where $\xi=\sigma\dfrac{\theta}{\sqrt{1+\theta^2}}\sqrt{\dfrac{2}{\pi}}$.
Without loss of generality, we can characterize the excursion of $\theta \equiv \theta(x)$ in the transition region by its maximum value $\bar{\theta}$, positive in the F$\to$C direction, and negative in the C$\to$F direction, so that $p(s|x) \equiv f(s|\mu,\sigma(x),\theta(x))$.
For a visual representation of the evolution of the jump-length skew-normal pdfs across the transition region, see Fig.~\ref{fig:skewed_pdfs}.

The skewness of $p(s|x)$ enters the expression for the mean, and thus directly induces a drift velocity correction localized at the interface, $v'=\xi/\tau$, which can be expressed as
\begin{equation}\label{eq:dv}
     v' =  \pm 2 \lambda \bar{\theta} \sqrt{\frac{v \, \alpha(1/2)}{\pi (1+\bar{\theta}^2)}} \delta(x - \frac{1}{2})
\end{equation}
where $\pm$ signs correspond to the F$\to$C and C$\to$F directions, respectively, and $\lambda=1$ has dimensions of length over square root of time.

\begin{figure}[t]
\centerline{\epsfxsize=9.0cm\epsfbox{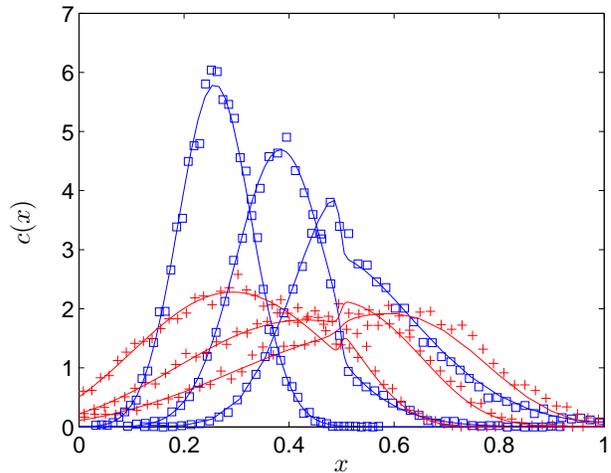}}
\caption{Resident concentration profiles corresponding to times $t=0.5$, $0.75$ and $1$, at $v=0.5$. Dispersivities are $\alpha_F=0.01$ and $\alpha_C=0.08$, and $\bar{\theta}=0.35$. Monte Carlo simulation results are displayed as symbols: squares correspond to F$\to$C flow conditions, crosses to C$\to$F. Comparisons with the rectified flux model are shown as solid lines.}
   \label{fig:res_drift}
\end{figure}

We can interpret this rectification flux as arising from subtle ratchet-like mechanisms at work in the interface between the two layers, $\mathcal{I}$. In other words, small perturbations in the potential energy of the particles in $\mathcal{I}$ do not simply average out, but rather induce a succession of asymmetric potential wells where the particles can jump with a preferential direction. Note that at high velocities the particles have enough energy to overcome the potential barriers, i.e., the advective part of the solute flux dominates over the dispersive part, and the differences between the BTCs in the two directions disappear. A variety of rectification fluxes have been experimentally observed in other systems where nonequilibrium fluctuations (endogeneous and/or exogeneous) in anisotropic media induce a unidirectional bias in the Brownian motion of a particle (see, e.g.,~\cite[]{kehr, astumian} and refs.~therein).

In Fig.~\ref{fig:btc_drift}, we display one-dimensional Monte Carlo random walk simulations with skew-normal pdfs $p(s)=f(s)$ for the breakthrough curves (BTC) measured at the exit of the column. Three sets of F$\to$C (squares) and C$\to$F curves (crosses) are shown, for increasing values of the velocity $v$. Our simulations qualitatively reproduce the salient features experimentally observed in~\cite{cortis_wrr}: in particular, $(i)$ the C$\to$F BTC is delayed with respect to the F$\to$C BTC; $(ii)$ as $v$ increases, the BTC curves become progressively closer to each other (i.e., the delay vanishes), and the standard AD behavior is recovered.

Monte Carlo estimates of the velocity correction at the interface, $v'$, are in good agreement with the predictions in Eq.~\eqref{eq:dv}. Our model is robust to changes in the functional form of the $p(s|x)$, as long as the second moment is finite and a spatial variation of the skewness is preserved. %
%


Concentration profiles along the column's longitudinal direction provide interesting clues about the nature of this transport process.
The concentration profiles in Fig.~\ref{fig:res_drift} display significant mass accumulation and sharp gradients at the interface, two characteristic signatures of our model. A non-invasive concentration profile laboratory measurement may be thus designed to validate the physical mechanism proposed in this work.

The expression for the solute flux can be written as $j^{\prime}(x,t) \equiv j^{AD}(x,t) +  v^{\prime}(x) c(x,t)$, which, owing to an integration by parts argument and neglecting the term $\partial_y \delta(y - \frac{1}{2})$, can also be recast as $j^{\prime}(x,t) \equiv j^{AD}(x,t) + \int_0^x v^{\prime}(y) \partial_y c(y,t)\, dy$. The equivalence between these two flux expressions shows how the interface correction is neither purely advective, nor purely dispersive, as also apparent from our random walk BTCs. We can now numerically solve the partial differential equation (pde) resulting by inserting the expression for $j'(x,t)$ into the continuity equation, and compare the results with the Monte Carlo simulations. This is done in Fig.~\ref{fig:btc_drift} for the BTCs and in Fig.~\ref{fig:res_drift} for the concentration profiles. The overall agreement between the Monte Carlo and pde approaches is excellent. From Eq.~\eqref{eq:dv}, we note that the rectified flux depends on the direction of the flow, being positive in the F$\to$C direction and negative in the opposite C$\to$F direction. The magnitude of the correction flux is small when dispersivity contrast is small, and/or the velocity $v$ is high. Also, the correction is smaller where the concentration gradient is smaller, i.e., at saturation. Since the thickness of the physical interface between the two layers strongly depends on the fine details of the Stokes velocity field, we expect $\bar{\theta}$ to increase with the inverse of the velocity $v$, proportionally to the dispersivity and permeability contrast.

When the velocity $v$ decreases below some given threshold, diffusion dominates over dispersion: the asymmetric transport mechanism can readily be adapted to describe this situation. Depending on the specific values of the diffusion coefficients for the two homogeneous sections, our model predicts that the separation between BTCs could even undergo an inversion, i.e., the F$\to$C curve could be delayed with respect to the C$\to$F. This is a pertinent issue, e.g., for nuclear waste storage in multilayered geological formations, where diffusion is (almost everywhere) the dominant transport mechanism. Differently from \cite{hornung}, our model preserves the continuity of the resident concentration at the interface.

Finally, we can extend our model to the case in which the two sections are characterized by a $\psi(t)$ pdf with slower than exponential decay (small degree of disorder)~\cite[]{cortis_chen}. Owing to the uncoupling between $p(s)$ and $\psi(t)$, the asymmetric transport mechanism at the interface remains unchanged, and the resulting pde will be defined through the time-convolution of $j'(x,t)$ with a memory function kernel depending only on $\psi(t)$~\cite[]{rev_geo}.

\acknowledgments
This work was supported, in part, by the U.S. Department of Energy under Contract No.~\mbox{DE-AC02-05CH11231}. We thank B.~Berkowitz, H.~Scher, J.G.~Berryman, Ph.~Montarnal, and the anonymous reviewers for interesting scientific discussions and constructive criticism.

\end{document}